# Efficient and Verifiable Privacy-Preserving Convolutional Computation for CNN Inference with Untrusted Clouds

Jinyu Lu[1], Xinrong Sun[1], Yunting Tao[2], Tong Ji[1], Fanyu Kong[1,3(✉)], Guoqiang Yang[3]

[1] School of Software, Shandong University, Jinan 250101, Shandong, China
[2] College of Information Engineering, Binzhou Polytechnic, Binzhou, 256603, Shandong, China
[3] Shandong Sansec Information Technology Co., Ltd, Jinan 250101, Shandong, China
`fanyukong@sdu.edu.cn`

**Abstract.** The widespread adoption of convolutional neural networks (CNNs) in resource-constrained scenarios has driven the development of Machine Learning as a Service (MLaaS) system. However, this approach is susceptible to privacy leakage, as the data sent from the client to the untrusted cloud server often contains sensitive information. Existing CNN privacy-preserving schemes, while effective in ensuring data confidentiality through homomorphic encryption and secret sharing, face efficiency bottlenecks, particularly in convolution operations. In this paper, we propose a novel verifiable privacy-preserving scheme tailored for CNN convolutional layers. Our scheme enables efficient encryption and decryption, allowing resource-constrained clients to securely offload computations to the untrusted cloud server. Additionally, we present a verification mechanism capable of detecting the correctness of the results with a success probability of at least $1 - \frac{1}{|Z|}$. Extensive experiments conducted on 10 datasets and various CNN models demonstrate that our scheme achieves speedups ranging $26 \times \sim 87 \times$ compared to the original plaintext model while maintaining accuracy.

## 1 Introduction

With the continuous development and enhancement of convolutional neural networks (CNNs), these models have demonstrated exceptional performance in computer vision [1], particularly in the domains of image classification [2, 3], face recognition [4, 5], and medical image analysis [6, 7]. The training and inference of neural network models demand substantial computational resources, making it challenging for resource-constrained devices such as mobile devices and small servers to accommodate these tasks. To address this issue, researchers in recent years have proposed a Machine Learning as a Service (MLaaS) system [8] for cloud platforms. This approach reduces the barriers to client-side model training and deployment by hosting the training and inference of neural network models on the cloud server, but it presents challenges in protecting privacy.

To enable the cloud server to complete the inference, the client must send input data to the cloud server. In many fields, including business and healthcare, this data frequently contains sensitive information, such as trade secrets [9] and personal details [10, 11]. The cloud server is often perceived as an untrustworthy third party [12], so clients expect that the privacy of the data transmitted to the cloud server is protected, rather than being sent in plaintext. Consequently, researchers have proposed privacy-preserving solutions for CNNs using various methods, including homomorphic encryption and secret sharing.

In [13], Hesamifard et al. applied homomorphic encryption to convolutional neural networks, enabling a cloud server to perform inference tasks by executing polynomial computations on encrypted data. However, this approach suffers from reduced accuracy and inefficiency. In [14], Riazi et al. attempted to address these issues with a hybrid secure computing framework; however, their method resulted in considerable communication overhead. In [15], Wagh et al. employed a secret sharing technique that divides the data into multiple copies, with the server retaining only one. This approach enhances both security and efficiency. Subsequently, both Juvekar et al. [16] and Zhang et al. [17] optimized linear computations in privacy-preserving CNNs to improve computational efficiency. Recently, in [18], Lee et al. proposed a residual number system FHE scheme that enhances existing homomorphic encryption schemes by extending their application to more general CNN models while achieving further optimizations in accuracy.

Despite these advancements, existing CNN privacy-preserving schemes still have considerable room for improvement in terms of efficiency. In state-of-the-art privacy-preserving CNN schemes, convolutional operations typically dominate the total computation time [17]. Consequently, there is significant potential for efficiency improvements by optimizing the handling of these convolutional computations. Based on the observations presented above, we propose a novel privacy-preserving scheme for convolutional layers. Our contributions are as follows:

- We propose an efficient, verifiable, and privacy-preserving scheme for the convolutional layer of CNNs. This approach allows resource-constrained clients to perform inference tasks by efficiently encrypting and decrypting data and securely delegating computational tasks to the cloud server.
- In response to the malicious behavior of the cloud server, our scheme provides an effective verification method that allows the client to detect errors when a cloud server returns incorrect computation results, with a probability of at least $1 - \frac{1}{|Z|}$.
- Through extensive benchmarking of various state-of-the-art CNN models on 10 different datasets, we demonstrate that our scheme achieves speedups ranging $26 \times \sim 87 \times$ compared to the original plaintext model while maintaining a comparable level of accuracy.

## 2 System Design

### 2.1 System Model

Our privacy-preserving scheme consists of two entities, including the client and the cloud server. Fig. **1** illustrates the interaction between these two entities within the privacy-preserving CNN framework.

1. Client $\mathcal{C}$: The $\mathcal{C}$ is a resource-constrained end device with private data. Within the convolutional layer, it obtains the model convolutional kernel data W from the $\mathcal{S}$ and sends input data and computational tasks to the $\mathcal{S}$.
2. Cloud server $\mathcal{S}$: The $\mathcal{S}$ is equipped with sufficient computational resources, hosting and exposing a well-trained CNN model that provides inference services to the $\mathcal{C}$. However, there is a risk that the $\mathcal{S}$ may be malicious, potentially stealing the $\mathcal{C}$'s private data or returning incorrect computational results.

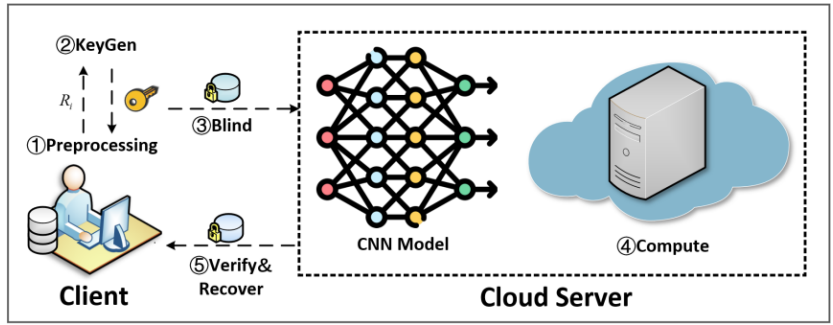

**Fig. 1.** System Model.

### 2.2 Framework

A privacy-preserving scheme consists of the following five algorithms:

- $SK \leftarrow KeyGen(\lambda, x)$: Given a security parameter $\lambda$, the $KeyGen$ algorithm returns a key $SK$, which is used by the $\mathcal{C}$ utilizes to blind the data.
- $(\delta_x, \vartheta_x) \leftarrow Blind(SK, x)$: The $\mathcal{C}$ blinds the data $x$ using $SK$ and returns a tuple $(\delta_x, \vartheta_x)$, where $\delta_x$ represents the blinded data and $\vartheta_x$ is employed for verification in the $Verify$ algorithm.
- $\delta_y \leftarrow Compute(\delta_x, F)$: $\mathcal{S}$ conducts the computation task $F$ on the data $\delta_x$ and returns the result $\delta_y$.
- $v \leftarrow Verify(\delta_y, \vartheta_x)$: The $Verify$ algorithm is utilized to verify the correctness of the results computed by the $\mathcal{S}$. If the result is deemed correct, the return value is $v = 1$; otherwise, it is $v = 0$.
- $\{y, \bot\} \leftarrow Recover(SK, \delta_y, v)$: If $v = 1$, the $\mathcal{C}$ uses $SK$ to recover the computational result and obtain $y$; otherwise, it returns $\bot$.

## 2.3 Design Goals

To ensure the viability of our privacy-preserving scheme, the following requirements must be met:

1. *Correctness*: The $\mathcal{S}$ within our architecture should be capable of accurately performing convolutional computation on encrypted image data. The results of the blinded computations can subsequently be decrypted by the $\mathcal{C}$ to perform subsequent tasks.
2. *Privacy Preservation*: Our scheme should prevent the $\mathcal{S}$ from obtaining any content or extracting features from the image. Even if $\mathcal{S}$ is aware of our privacy-preserving scheme, it will be unable to extract any useful information from the data provided by the $\mathcal{C}$, including both input and output information.
3. *Verifiability*: In our architecture, *the $\mathcal{S}$* is considered a malicious third party. Therefore, our scheme must ensure that if the $\mathcal{S}$ returns a correct computation result, the $\mathcal{C}$ can detect with probability 1 that the result is correct. Conversely, if $\mathcal{S}$ returns an incorrect result, the $\mathcal{C}$ should be able to identify with a high probability that the result is incorrect.
4. *Efficiency*: The cost to the $\mathcal{C}$ for executing the scheme, which includes blind, verification, and recovery, should be lower than the cost of executing the original computation.

## 3 Our Proposed Scheme

In this section, we introduce our proposed privacy-preserving scheme for CNNs. Firstly, Fig. **2** illustrates the details of our scheme, and Table **1** provides explanations of the notations used in this paper.

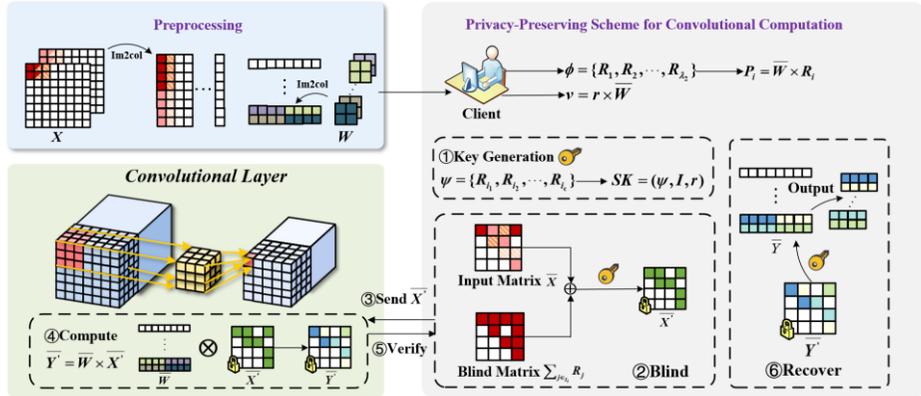

**Fig. 2.** Privacy-preserving scheme for CNNs.

**Table 1.** The explanation of notations

| Notation | Explanation |
| --- | --- |
| $X$ | Input matrix |
| $Y$ | Output matrix |
| $W$ | Convolution kernel |
| $m$ | The length of the input matrix |
| $n$ | The width of the input matrix |
| $k$ | The size of the convolution kernel |
| $c_{in}$ | Number of input channels |
| $c_{out}$ | Number of output channels |
| $v$ | Verification vector |
| $\mathcal{C}$ | Client |
| $\mathcal{S}$ | Cloud server |
| $SK$ | Secret key |
| $\lambda$ | Security parameter |

### 3.1 Preprocessing

Consider the example of a single convolution operation of an input image within a convolutional layer. For simplicity, we assume the stride parameter is 1, which is the most common configuration. We analyze the computational task $Conv(X, W, 1)$, which applies the convolution kernel $W$ to the input image $X$ with a stride size of 1. The input dimensions are $(m, n, c_{in})$, the output dimensions are $(k, c_{in}, c_{out})$, and the size of the convolutional kernel is $k$.

Due to the number of input channels and output channels $(c_{in}, c_{out})$, the total number of convolution kernels is $c_{in} \times c_{out}$. The set of convolution kernels is represented as:

$$W = \left\{\left(W_1^{(1)}, W_1^{(2)}, \dots, W_1^{(c_{in})}\right), \left(W_2^{(1)}, W_2^{(2)} \dots, W_2^{(c_{in})}\right), \dots, \left(W_{c_{out}}^{(1)}, W_{c_{out}}^{(2)}, \dots, W_{c_{out}}^{(c_{in})}\right)\right\}$$

For an image, the set of its corresponding input matrices is represented as:

$$X = \left(X^{(1)}, X^{(2)}, \dots, X^{(c_{in})}\right)$$

The $\mathcal{C}$ performs the following actions in sequence.

1. **Constructing the Im2col Matrix:** For each input $X^{(j)}$, construct a series of submatrices as follows:

$$Sub(X^{(j)})_{p,q} = \begin{bmatrix} x_{p,q} & \cdots & x_{p,q+k-1} \\ \vdots & \ddots & \vdots \\ x_{p+k-1,q} & \cdots & x_{p+k-1,q+k-1} \end{bmatrix}, 1 \leq j \leq c_{in}, \quad (1)$$

where $x_{p,q}$ denotes the element in the $p$-th row and $q$-th column of the input matrix $X^{(j)}$, with the conditions $1 \leq p \leq (m-k+1)$ and $1 \leq q \leq (n-k+1)$.

Each input matrix $X^{(j)}$ will construct $(m-k+1)(n-k+1)$ submatrices, which are subsequently transformed into column vectors as follows:

$$Sub(X^{(j)})_{p,q} = \begin{bmatrix} x_{p,q} & \cdots & x_{p,q+k-1} & \cdots & x_{p+k-1,q} & \cdots & x_{p+k-1,q+k-1} \end{bmatrix}^T, \quad (2)$$

where $s_{i,j}$ denotes the element located in the $i$-th row and $j$-th column of matrix $Sub(X^{(j)})_{p,q}$. The im2col matrix is then constructed according to the following rules:

$$\overline{X} = \begin{bmatrix} Sub(X^{(1)})_{1,1} & \cdots & Sub(X^{(1)})_{1,n-k+1} & Sub(X^{(1)})_{2,1} & \cdots & Sub(X^{(1)})_{m-k+1,n-k+1} \\ Sub(X^{(2)})_{1,1} & \cdots & Sub(X^{(2)})_{1,n-k+1} & Sub(X^{(2)})_{2,1} & \cdots & Sub(X^{(2)})_{m-k+1,n-k+1} \\ \vdots & \ddots & \vdots & \vdots & \ddots & \vdots \\ Sub(X^{(c_i)})_{1,1} & \cdots & Sub(X^{(c_{in})})_{1,n-k+1} & Sub(X^{(c_{in})})_{2,1} & \cdots & Sub(X^{(c_{in})})_{m-k+1,n-k+1} \end{bmatrix}. \quad (3)$$

2. **Constructing the Kernel Matrix:** Each convolution kernel is transformed into a row vector as follows:

$$W_i^{(j)} = \begin{bmatrix} w_{1,1} & \cdots & w_{1,k} & \cdots & w_{k,1} & \cdots & w_{k,k} \end{bmatrix}, 1 \leq j \leq c_{in}, \quad (4)$$

where $w_{i,j}$ denotes the element located in the $i$-th row and $j$-th column of matrix $W_i^{(j)}$. These row vectors are subsequently assembled into a kernel matrix as follows:

$$\overline{W} = \begin{bmatrix} W_1^{(1)} & W_1^{(2)} & \cdots & W_1^{(c_{in})} \\ W_2^{(1)} & W_2^{(2)} & \cdots & W_2^{(c_{in})} \\ \vdots & \vdots & \ddots & \vdots \\ W_{c_{out}}^{(1)} & W_{c_{out}}^{(2)} & \cdots & W_{c_{out}}^{(c_{in})} \end{bmatrix}. \quad (5)$$

3. **Precomputation:** Before the inference task begins, the $\mathcal{C}$ randomly generates the set $\Phi = \{R_1, R_2, \ldots, R_{\lambda_2}\}$, which contains $\lambda_2$ matrices, each with dimensions $c_i k^2 \times (m-k+1)(n-k+1)$, with elements in the matrices taking the range $(-2^{\lambda_1}, 0)$ and $(0, 2^{\lambda_1})$, where $\lambda_1$ and $\lambda_2$ is the security parameters. The $\mathcal{C}$ then computes:

$$P_l = \overline{W} \times R_l, 1 \leq l \leq \lambda_2. \quad (6)$$

The $\mathcal{C}$ securely stores the matrix sets $\{R_1, R_2, \ldots, R_{\lambda_2}\}$ and $\{P_1, P_2, \ldots, P_{\lambda_2}\}$ for subsequent blinding and recovery.

### 3.2 Privacy-Preserving Convolutional Computation

The following operations are performed for each matrix $\overline{X}$:

1. **Key Generation:** The $\mathcal{C}$ randomly selects matrices from the set $\phi$ to create the set $\psi = \{R_{i_1}, R_{i_2}, \ldots, R_{i_c}\}(\psi \subseteq \phi, \psi \neq \emptyset)$. It then generates index $I = \{i_1, i_2, \ldots, i_c\}$, and generates a $c_{out}$ dimensional random row vector $r$, with each element of the vector taking values in the range $(-2^{\lambda_1}, 0)$ and $(0, 2^{\lambda_1})$, to form the $SK = (\psi, I, r)$.

2. **Blind:** The $\mathcal{C}$ blinds the input matrix by sequentially adding the matrices from the set $\psi$ to the input matrix $\overline{X}$, as demonstrated in the following equation:

$$\overline{X'} = \overline{X} + \sum_{l \in I_i} R_l. \tag{7}$$

Next, the $\mathcal{C}$ generates a validation vector $v$ utilizing the following equation:

$$v = r \times \overline{W}, \tag{8}$$

which assists in verifying the results returned by the $\mathcal{S}$ during the verification phase. The $\mathcal{C}$ then sends the blinded matrix $\overline{X'}$ to the $\mathcal{S}$.

3. **Compute:** After the $\mathcal{S}$ receives the data $\overline{X'}$ from the $\mathcal{C}$, it computes the data according to the following equation:

$$\overline{Y'} = \overline{W} \times \overline{X'}, \tag{9}$$

and returns the result $\overline{Y'}$ to the $\mathcal{C}$.

4. **Verify:** After receiving the result from the $\mathcal{S}$, the $\mathcal{C}$ first verifies the computed result using the following equation:

$$r \times \overline{Y'} = V \times \overline{X'}, \tag{10}$$

where $r$ represents the random vector in the key $SK$, and $v$ denotes the verification vector generated during the blinding phase. If this equation is satisfied, the verification is considered successful; otherwise, it is deemed unsuccessful and this result is discarded.

5. **Recover:** The $\mathcal{C}$ deblinds the matrix $\overline{Y'}$ by sequentially subtracting the matrix $P_l$ corresponding to index $I$ as demonstrated in the following equation:

$$\overline{Y} = \overline{Y'} - \sum_{l \in I_i} P_l, \tag{11}$$

where $P_j$ is the matrix generated by the $\mathcal{C}$ in the precomputation phase.

The matrix $\overline{Y}$ has dimensions $c_{out} \times (m-k+1)(n-k+1)$. As shown in the following equation, it is partitioned into several smaller matrices, each denoted as $y_b^{(a)}$, with dimensions $1 \times (n-k+1)$:

$$\overline{Y} = \begin{bmatrix} y_1^{(1)} & y_2^{(1)} & \cdots & y_{m-k+1}^{(1)} \\ y_1^{(2)} & y_2^{(2)} & \cdots & y_{m-k+1}^{(2)} \\ \vdots & \vdots & \ddots & \vdots \\ y_1^{(c_{out})} & y_2^{(c_{out})} & \cdots & y_{m-k+1}^{(c_{out})} \end{bmatrix}. \tag{12}$$

The output is subsequently derived by reconstructing the matrix using the following equations:

$$Y^{(t)} = \begin{bmatrix} y_1^{(t)} & y_2^{(t)} & \cdots & y_{m-k+1}^{(t)} \end{bmatrix}^T, 1 \leq t \leq c_{out}. \quad (13)$$

After completing the recovery operation, the output matrices set $Y = (Y^{(1)}, Y^{(2)}, \ldots, Y^{(c_{out})})$ is utilized as the input for the next layer to continue with the subsequent tasks.

## 4 Analysis

In this section, we analyze our scheme in terms of correctness, security, verifiability, and efficiency.

### 4.1 Correctness

**Theorem 1.** *Provided that the $\mathcal{C}$ and $\mathcal{S}$ accurately execute the computations according to our scheme, the resulting outputs will be correct.*

*Proof.* The $\mathcal{C}$ blinds and de-blinds the data utilizing *eq.* (7) and *eq.* (11), respectively, while the $\mathcal{S}$ performs the computational task and returns the result to the $\mathcal{C}$ according to *eq.* (9). Assuming that both the $\mathcal{C}$ and the $\mathcal{S}$ follow our scheme, the following equation holds:

$$\overline{Y} + \sum_{j \in I} P_j = \overline{W} \times \left( \overline{X} + \sum_{j \in I} R_j \right). \quad (14)$$

According to the precomputation conducted in *eq.* (6), we have:

$$\overline{Y} = \overline{W} \times \overline{X} + \left( \overline{W} \times \sum_{j \in I} R_j - \sum_{j \in I} P_j \right) = \overline{W} \times \overline{X}. \quad (15)$$

Therefore, for any convolutional kernels $W$, input matrix $X$, and de-blinded matrix $\overline{Y_i}$, the following equation holds: $\overline{Y} = \overline{W} \times \overline{X}$. Thus, the resulting outputs are correct.

### 4.2 Privacy Preservation

**Theorem 2.** *Given arbitrary data, even if $\mathcal{S}$ is aware of our privacy-preserving scheme, our scheme is still capable of safeguarding both the input and output privacy of the data while ensuring that no information is disclosed.*

*Proof.* Consider a single convolution operation. Let $Z = (-2^{\lambda_1}, 0) \cup (0, 2^{\lambda_1})$, input data $X$, the convolution kernel $W$, and $SK = (\psi, I, r)$. Let $B = \sum_{j \in I} R_j$, we denote the set of values for matrix B as $Z'$. It follows that $Z \subseteq Z'$.

Considering input privacy, the $\mathcal{S}$ can recover data by either guessing the $SK$ or the matrix $B$. Suppose the dimension of matrix $X$ is $n \times m$. Each matrix in the set $\psi$ has $|Z|^{nm}$ possible configurations, and there are $2^{\lambda_2} - 1$ possible combinations for the set $\psi$. Therefore, the size of the key space for $SK$ is $(2^{\lambda_2} - 1)|Z|^{nm}$. The probability that the $\mathcal{S}$ accurately guesses the $SK$ in a single attempt is $\frac{1}{(2^{\lambda_2}-1)|Z|^{nm}}$, which is extremely small and can be considered negligible. The dimensions of matrix $B$ is $n \times m$, resulting in a total of $|Z'|^{nm}$ possible configurations. Similarly, the probability that the $\mathcal{S}$ accurately guesses the $SK$ in a single attempt is $\frac{1}{|Z'|^{nm}}$, which is extremely small and can be considered negligible.

Considering output privacy, the $\mathcal{S}$ can either recover the input data first and then compute the output data using the key, or compute $W \times B$ to recover the output data. This means that the $\mathcal{S}$ must also guess the $SK$ or matrix $B$, making this probability negligible as well. Therefore, our scheme protects the privacy of data input and output, while ensuring that no information is disclosed to $\mathcal{S}$.

For client-server communication, data privacy is achieved through the encryption of transmitted information. Even if the third party intercepts the communication, they would be unable to extract any meaningful information.

### 4.3 Verifiability

**Theorem 3.** Our privacy-preserving scheme is verifiable, with the probability of an incorrect verification judgment is less than or equal to $\frac{1}{|Z|}$.

*Proof.* Consider a single convolution operation and assuming the $\mathcal{S}$ returns an incorrect result $\overline{Y'}$. According to *eq.* (8) executed by the $\mathcal{C}$ in the blinding phase and *eq.* (10) executed in the verification phase, if the verification is successful, the following equation is satisfied:

$$r \times \overline{Y'} = r \times (\overline{W} \times \overline{X'}). \tag{16}$$

Let $M = \overline{Y'} - \overline{W} \times \overline{X'}$, the equation above transforms into:

$$r \times M = 0. \tag{17}$$

For any $s$, $1 \leq s \leq (m-k+1)(n-k+1)$, it holds that $\sum_{k=1}^{c_{out}} r_k m_{ks} = 0$. Since the result is incorrect, it follows that $M \neq 0$. Therefore, there exists an element $m_{ij} \neq 0$. We consider the $j$-th column of $M$. Let $a_k = m_{kj}$, the original equation can be interpreted as solving the linear equation for $r$:

$$a_1 r_1 + a_2 r_2 + \cdots + a_{c_{out}} r_{c_{out}} = 0, \tag{18}$$

where $a_j \neq 0$. Thus, the equation above can be rewritten as:

$$r_j = -\frac{1}{a_j} \sum_{i \neq j}^{c_{out}} a_i r_i. \tag{19}$$

Let $S = -\frac{1}{a_j} \sum_{i \neq j}^{c_{out}} a_i r_i$, when $S \notin Z$, $r_j$ has no solution; conversely, when $S \in Z$, $r_j$ has a solution and $\Pr(r_j = S) = \frac{1}{|Z|}$. According to the Total Probability Theorem, we have:

$$\Pr(r \times M = 0 | M \neq 0) = \Pr(\sum_{k=1}^{c_{out}} r_k m_{kj} = 0 | S \notin Z) \Pr(S \notin Z)$$

$$+\Pr\left(\sum_{k=1}^{c_{out}} r_k\, m_{kj} = 0 \big| S \in Z\right)\Pr(S \in Z)$$
$$= 0 + \frac{1}{|Z|}\Pr(k \neq 0) \leq \frac{1}{|Z|}. \tag{20}$$

Considering that $|Z|$ is a very large number, the probability of the result passing the $\mathcal{C}$'s verification is negligible if the $\mathcal{S}$ returns an incorrect result. Therefore, our scheme is verifiable.

### 4.4 Efficiency

In our privacy-preserving scheme, the $\mathcal{C}$ is primarily responsible for the computations associated with the blinding, verification, and recovery phases, while the $\mathcal{S}$ is primarily responsible for matrix multiplication. In the following efficiency analysis, we examine a single convolution operation involving an input matrix, specifically focusing on the computational overhead associated with scalar multiplication, denoted by $SM$, and scalar addition, denoted by $SA$. The computational overhead is detailed in Table **2**

For the $\mathcal{C}$, in the blinding phase, the primary overhead consists of computing *eq.* (7) and *eq.* (8), which involves executing $|I_i|c_{in}k^2(m-k+1)(n-k+1)$ $SAs$ and $c_{in}c_{out}k^2$ $SMs$. In the verification phase, the primary overhead consists of computing *eq.* (10), which involves executing $(c_{out}+c_{in}k^2)(m-k+1)(n-k+1)SMs$. In the recovery phase, the primary overhead consists of computing *eq.*(11), which involves executing $|I_i|c_{out}(m-k+1)(n-k+1)$ $SAs$. Therefore, the total computational overhead for the $\mathcal{C}$ is $[c_{in}c_{out}k^2 + (c_{out}+c_{in}k^2)(m-k+1)(n-k+1)]SMs + |I_i|(c_{out}+c_{in}k^2)(m-k+1)(n-k+1)SAs$. For the $\mathcal{S}$, overhead consists of computing *eq.* (9), which involves executing $c_{in}c_{out}k^2(m-k+1)(n-k+1)$ $SMs$.

Considering that the convolution kernel size $k$ is typically much smaller than the image dimensions $m$ and $n$, where $m$ and $n$ are generally larger values. It is important to note that in computer floating-point arithmetic, scalar addition is often regarded as a lightweight operation compared to scalar multiplication. Therefore, the ratio of $\mathcal{C}$ to $\mathcal{S}$ overhead can be approximated as:

$$\frac{c_{out}+c_{in}k^2}{c_{in}c_{out}k^2}, \tag{21}$$

which approaches zero. In other words, our scheme significantly reduces the overhead on the $\mathcal{C}$ side when performing convolutional computations, thereby demonstrating its efficiency.

**Table 2.** Computational overhead

|  | SM | SA |
|---|---|---|
| Blind | $c_{in}c_{out}k^2$ | $|I_i|c_{in}k^2(m-k+1)(n-k+1)$ |
| Verify | $(c_{out}+c_{in}k^2)(m-k+1)(n-k+1)$ | 0 |
| Recover | 0 | $|I_i|c_{out}(m-k+1)(n-k+1)$ |

## 5 Experiments

### 5.1 Experimental Settings

We developed a prototype that utilizes two computers to simulate a $\mathcal{C}$ and a $\mathcal{S}$, employing Python programming. The $\mathcal{C}$ is equipped with an Intel Core i7 12700H CPU and 16 GB of RAM, while the $\mathcal{S}$ features an Intel Xeon CPU, 32 GB of RAM, and an NVIDIA V100 GPU with CUDA acceleration. It is important to emphasize that our experiment is a simulation and that real-world device resources may vary. Nevertheless, the $\mathcal{C}$ without a GPU in our simulation is representative of typical resource-constrained devices.

To systematically evaluate our scheme under various conditions, we generated a series of simulated datasets, each consisting of 100 randomly generated multichannel noisy images. These datasets are labeled as TEST-$i$, where $i \in \{3, 50, 100, \ldots, 300\}$ indicates the number of input channels. To rigorously assess the effectiveness of our scheme across different visual modalities and models, we benchmarked it on 10 real-world datasets, utilizing 7 state-of-the-art CNN architectures. Specifically, for grayscale image datasets, we employed the MNIST[1] and Fashion-MNIST[2] datasets and utilized the VGG-16 [3] and VGG-19 [3] models. For RGB image datasets, we conducted experiments on the CIFAR-10, IFAR-100[3], PASCAL VOC[4], Cityscapes[5], and Stanford Dogs[6] datasets, using the ResNet-50 [19], ResNet-101 [19], and ResNet-152 [19] models. Lastly, for hyperspectral image datasets, we evaluated the Pavia University, Indian Pines, and KSC datasets[7] using 2D-CNN[20] and 3D-CNN [20] models with three convolutional layers.

### 5.2 Evaluation Result

We initially evaluate the efficiency of our scheme using simulated datasets. We conducted inference with a three-layer convolutional 3D-CNN model using the TEST-50, TEST-100, …, and TEST-300 datasets. Additionally, we evaluated different output channel sizes individually using the TEST-3 dataset. We evaluated the time cost associated with each phase of execution, where the blinding, verification, and recovery phases are performed on the $\mathcal{C}$, while the computation phase is executed on the $\mathcal{S}$.

---

[1] *MNIST handwritten digit dataset* [Online]. Available: http://yann.lecun.com/exdb/mnist
[2] *Fashion-MNIST dataset.* Download: https://dblp.org/rec/bib/journals/corr/abs-1708-07747
[3] *CIFAR-10, CIFAR-100 datasets.* [Online]. Available: https://www.cs.toronto.edu/~kriz/cifar.html
[4] *PASCAL VOC dataset.* [Online]. Available: http://host.robots.ox.ac.uk/pascal/VOC/
[5] *Cityscapes dataset.* [Online]. Available: https://www.cityscapes-dataset.com/
[6] *Stanford Dogs dataset.* [Online]. Available: http://vision.stanford.edu/aditya86/ImageNetDogs/main.html
[7] *Pavia University, Indian Pines, and KSC datasets.* [Online]. Available: https://www.ehu.eus/ccwintco/index.php/Hyperspectral_Remote_Sensing_Scenes#Pavia_Centre_and_University%23Pavia_Centre_and_University

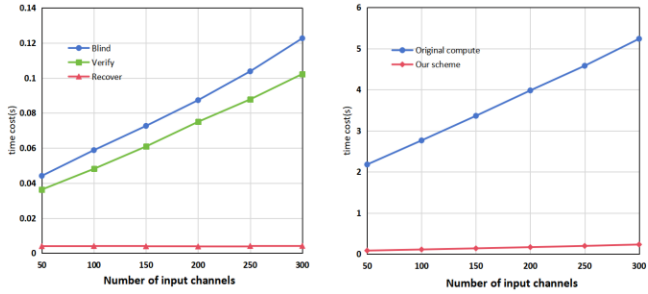

**(a)** Time cost in different phases   **(b)** Compare the time cost of our scheme with the original computation

**Fig. 3.** Time cost with different input channels.

Fig. **3** illustrates the average time cost required by our scheme to analyze each sample in the simulated dataset, based on different input channel sizes. Overall, the time cost for inference increases with the number of input channels. Specifically, Fig. **3 (a)** illustrates a comparison of the time cost for our scheme during the blinding, verification, and recovery phases. We observed that the time cost associated with the blinding phase is slightly higher than that of the verification phase, whereas the time cost for the recovery phase is significantly lower than both. The blinding phase includes both multiplication and addition operations, while the verification phase consists solely of multiplication operations. In contrast, the recovery phase involves only lightweight addition operations. These findings align with our analysis in Section 5.4.

Fig. **3(b)** illustrates a comparison of the time cost associated with the original computation and our proposed scheme. The original computation refers to executing the convolutional task on the $\mathcal{C}$ side without a privacy-preserving scheme. We observe that the time cost for the original computation is significantly higher than that of our proposed scheme, and this disparity becomes increasingly evident as the number of input channels increases. As a result, our scheme effectively processes images with a wide range of input channel counts, from standard images to hyperspectral images.

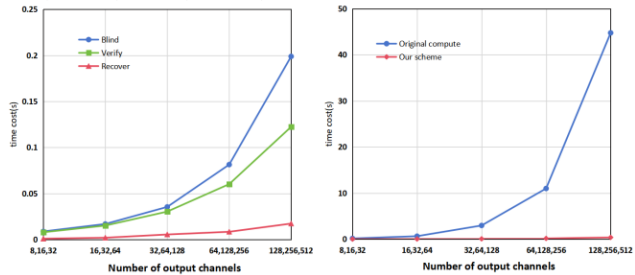

**(a)** Time cost in different phases   **(b)** Compare the time cost of our scheme with the original computation

**Fig. 4.** Time cost with different output channels

Fig. **4** illustrates the average time cost required by our scheme to analyze each sample in the simulated dataset, based on the output channel size. Overall, the time cost for inference increases with the number of output channels. Increasing the number of output channels typically enhances a CNN's ability to extract more features from the image. Fig. **4(a)** shows that while the $\mathcal{C}$ time cost increases rapidly with the number of output channels, it remains within acceptable limits. In contrast, Fig. **4(b)** indicates that the time cost of the original computation rises dramatically as the number of output channels increases. Therefore, our scheme is well-suited for CNN with multiple output channels, enabling the efficient extraction of a substantial number of features from image.

We subsequently evaluate the efficiency and accuracy of our scheme utilizing real-world datasets and classical CNN models.

Fig. **5** illustrates the average time cost required by our scheme to analyze each sample in a real dataset based on CNN models of different model complexities. Table **3** provides detailed data, where the time cost of our scheme includes the time required for the $\mathcal{C}$ blinding, verification, and recovery phases, as well as the time taken for $\mathcal{S}$ computation. Fig. **5(a)** illustrates that the time costs at each phase of our scheme on real-world datasets align with the findings from the previous analysis. Specifically, the blinding phase incurs a slightly higher time cost than the verification phase, while the recovery phase exhibits a significantly lower time cost compared to the first two phases. Notably, even with the most complex ResNet-152 model, the $\mathcal{C}$'s time cost remains below 14 seconds. Fig. **5(b)** illustrates that, compared to the original plaintext scheme, our approach exhibits a significantly lower time cost when applied to real-world datasets, particularly with complex CNN models like the ResNet Architecture Family. *Eq.* 21 demonstrates that as the number of convolution kernels $k$, input channels $c_{in}$, and output channels $c_{out}$ increases, the efficiency improvement of our scheme becomes more significant. More complex CNN models generally have a greater number of output channels. Consequently, as the complexity of the CNN model increases, the efficiency improvement of our scheme is enhanced.

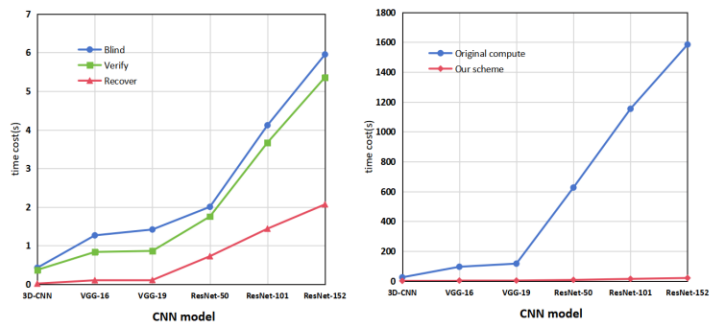

**(a)** Time cost in different phases     **(b)** Compare the time cost of our scheme with the original computation

**Fig. 5.** Time cost in different CNN models.

Table 3. Time cost with different models and datasets (ms)

| Models | Blind | Verify | Recover | Original compute | Our scheme | Speedup |
|---|---|---|---|---|---|---|
| 3D-CNN | 428.98 | 374.11 | 19.09 | 24655.81 | 925.38 | 26× |
| VGG-16 | 1268.05 | 838.17 | 105.6 | 95306.75 | 2571.38 | 37× |
| VGG-19 | 1419.22 | 865.32 | 110.77 | 116268.77 | 2884.62 | 40× |
| ResNet-50 | 2006.12 | 1756.10 | 727.93 | 626115.13 | 7125.07 | 87× |
| ResNet-101 | 4117.86 | 3670.04 | 1440.18 | 1153649.46 | 14083.04 | 81× |
| ResNet-152 | 5951.44 | 5359.54 | 2070.08 | 1584477.24 | 20049.09 | 79× |

Tables **4-6** provide a detailed comparison of the accuracy performance of our scheme versus the original plaintext scheme, utilizing various models and datasets. We observe that in most datasets, the accuracy of our scheme is nearly indistinguishable from that of the original plaintext scheme. This is due to the fact that our scheme does not introduce any approximations and does not lead to any theoretical loss of accuracy. There are only a few datasets where the accuracy of our scheme is slightly lower than that of the original plaintext scheme. The reason for this difference is that, in practice, the operation of floating-point numbers in the cumulative computation described in *eq.* (7) and *eq.* (11) may result in the accumulation of minor rounding errors, as well as potential sample selection bias. However, the loss of accuracy due to this error does not exceed 0.5%, which is considered negligible.

Table 4. Accuracy with VGG models and different datasets

| Dataset | VGG-16 | | VGG-19 | |
|---|---|---|---|---|
| | Original scheme (%) | Our scheme (%) | Original scheme (%) | Our scheme (%) |
| MNIST | 90.71 | 90.56 | 91.64 | 92.04 |
| Fashion-MNIST | 92.81 | 92.98 | 94.02 | 93.97 |

Table 5. Accuracy with ResNet models and different datasets

| Dataset | ResNet-101 | | ResNet-152 | |
|---|---|---|---|---|
| | Original scheme (%) | Our scheme (%) | Original scheme (%) | Our scheme (%) |
| CIFAR-10 | 93.79 | 93.74 | 94.49 | 94.22 |
| CIFAR-100 | 76.62 | 76.61 | 77.38 | 76.93 |
| PASCAL VOC | 94.42 | 94.34 | 94.98 | 94.53 |
| Cityscapes | 95.34 | 95.33 | 95.35 | 95.46 |
| Stanford Dog | 74.89 | 74.87 | 77.98 | 77.49 |

Table 6. Accuracy with 2D-CNN and 3D-CNN models and different datasets

| Dataset | 2D-CNN | | 3D-CNN | |
|---|---|---|---|---|
| | Original scheme (%) | Our scheme (%) | Original scheme (%) | Our scheme (%) |
| Pavia University | 93.70 | 93.95 | 99.86 | 99.86 |
| Indian Pines | 90.21 | 89.97 | 97.73 | 97.63 |
| KSC | 94.57 | 94.07 | 96.66 | 96.22 |

## 6  Conclusion

In this paper, we propose a novel verifiable privacy-preserving scheme tailored for CNN convolutional layers. Theoretical analysis demonstrates that our scheme achieves efficient encryption and decryption, allows resource-constrained clients to securely offload computations to the untrusted cloud server, and has the capability to detect malicious behavior in the cloud server with a probability of at least $1 - \frac{1}{|Z|}$. Experimental analysis demonstrates that our scheme achieves a speedup of $26 \times$ to $87 \times$ compared to the original plaintext model, while maintaining accuracy. In the future, we will conduct a more thorough investigation into the development of CNN privacy preservation, with a focus on efficiency and generalization.

**Acknowledgements.** This work is supported by the Key Research and Development Program of Shandong Province, China (Grant No. 2022CXGC020102), the Science and Technology Small and Medium Enterprises (SMEs) Innovation Capacity Improvement Project of Shandong Province & Jinan City, China (Grant No. 2022TSGC2048), the Haiyou Famous Experts - Industry Leading Talent Innovation Team Project of Shandong Province Jinan City, China.